

\def\NL{\hfill\break}
\def\NI{\noindent}
\magnification=\magstep 1
\overfullrule=0pt
\hfuzz=16pt
\voffset=0.1 true in
\vsize=9.2 true in
\def\NP{\vfil\eject}
\baselineskip 20pt
\parskip 6pt
\hoffset=0.1 true in
\hsize=6.2 true in
\nopagenumbers
\pageno=1
\footline={\hfil -- {\folio} -- \hfil}

\hphantom{AA}

\hphantom{AA}

\centerline{\bf EXACT RESULTS FOR A THREE-BODY REACTION-DIFFUSION
SYSTEM}

\vskip 0.4in

\centerline{\bf Vladimir~Privman$^{a,b}$}

\vskip 0.2in

\item{$^a$}{\sl{Department of Physics, Theoretical Physics, University
of Oxford, \NL 1 Keble Road, Oxford \ OX1 3NP, UK}}

\item{$^b$}on leave of absence from {\sl{Department of Physics, \NL
Clarkson University, Potsdam, New York \ 13699--5820, USA}}

\vskip 0.4in

\NI {\bf PACS:}$\;$ 05.70.Ln, 68.10.Jy, 82.20.Mj

\vskip 0.4in

\centerline{\bf ABSTRACT}

A system of particles hopping on a line, singly or as merged pairs,
and annihilating in groups
of three on encounters, is solved exactly for certain symmetrical
initial
conditions. The functional form of the density is nearly identical to
that found in two-body annihilation, and both systems show
non-mean-field, $\sim 1/\sqrt{t}$ instead of $\sim 1/t$,
decrease of particle density for large times.

\NP

Diffusion limited reactions in low dimensions
have attracted much theoretical interest [1-11].
Indeed, fluctuation effects are most profound in low dimensions,
implying deviations from the mean-field rate-equation behavior.
Numerical, analytical, and in few cases exact results have been
obtained for two-body
reactions $A+A \to {\rm products}$, and $A+B \to {\rm products}$,
in $d=1$.
Particle input and production (fragmentation) have also been
allowed for in recent studies [4,8,12].

Exact solutions are available only for the simplest reactions.
Specifically, for the two-body annihilation,
$A+A \to {\rm inert}$, the most detailed exact results have been
obtained by methods also used for the Glauber-dynamics Ising models.
The particles $A$ are then interfaces on a
lattice dual to the Ising spin lattice [5,9,13]; the time evolution is
continuous. Exact results for cellular-automata type,
simultaneous-update, discrete time dynamics, have also been derived
[11]. For reactions $A+A \to A$, possibly with particle input and
fragmentation, continuum-limit results have been obtained by various
random-walk arguments [7,12], some of which are not
limited [7] to $d=1$.

In this work, exact results will be obtained for a certain three-body
reaction scheme to be defined shortly. The method of solution involves
consideration of \break $Z_k$-symmetric, $k$-state ``spins'' on a linear
lattice. Multistate, Potts-spin $d=1$ lattice
has been considered [14] as a model of
mixed pairwise annihilation and coagulation of interfaces between
the fully ordered ($T=0$) Potts phases; the limit $k=\infty$
was solved by random-walk methods [14]. The approach adopted here is
quite different: discrete-time and space cellular-automaton dynamics
will be defined, and the case $k=3$ will be shown to have certain
simplified properties allowing an exact solution.

It is convenient to use directly the roots-of-unity complex
representation of $Z_k$, the numbers $\exp \left( 2\pi i n /k \right)$,
with $n=0,1,2, \ldots, k-1$. Our ``spin'' variables, $\sigma_j
(t)$, take on these $k$ distinct
values. Here the time $t$ increases in integer
steps, $t=0,1,2, \ldots$. The spin variables are at even lattice sites
$j=0, \pm 2, \pm 4, \ldots$, for even times, and at odd lattice sites
for odd times. Let

$$ \sigma_j (t) = \exp \left[ 2 \pi i
 n_j (t) /k \right] \eqno(1) $$

\NI We identify the reacting-diffusing
particles as follows. On a dual lattice,
i.e., in the interstice between two neighboring spins at $j-1$ and
$j+1$, we calculate the number $\sigma_{j+1} / \sigma_{j-1} =
\sigma_{j+1} \sigma^*_{j-1}$, where star denotes complex conjugate.
The result is put in the form
$\exp \left( 2 \pi i
 N_j /k \right)$, where $N_j = \left( n_{j+1}-n_{j-1}, \,
{\rm mod} \, k \right)$,
is a number in the range $0,\ldots,k-1$. We then identify
the number of particles at $j$ (at time $t$) as $N_j (t)$
thus calculated. Note that the particles occupy the dual lattice sites
with even $j$ for odd times, and with odd $j$ for even times $t$.

Next, we define the reaction-diffusion dynamics. At each time step
$t \to t+1$, all the $N_j (t)$ particles at the dual-lattice site
$j$ hop {\it as a group\/} to the right, to site $j+1$ of the dual
lattice, or to the left, to site $j-1$, with equal probability.
Thus the resulting number of particles hopping to site $m$ can be in the
range $0,\ldots,2k-2$. More precisely, this number can be 0,
$N_{m-1}$, $N_{m+1}$, or the sum $N_{m-1}+N_{m+1}$. However, we now
add the reaction process of
instantaneous annihilation of $k$ particles, at all sites at which
there are more than $k$ particles present after hopping. Thus,
in those sites $m$ which received $N_{m-1}(t)+N_{m+1}(t)$ particles,
we define $N_m (t+1)$ as $\left( N_{m-1} (t)+N_{m+1} (t)
, \, {\rm mod} \, k \right)$.

These dynamics can be put in a convenient linear form in terms of the
\break $Z_k$-symmetric spins,

$$ \sigma_j (t+1) = \left[ 1 - \zeta_j (t+1) \right]
\sigma_{j-1} (t) + \zeta_j (t+1) \sigma_{j+1} (t) \eqno(2) $$

\noindent Here we assigned independent and uncorrelated random
variables $\zeta_m (t)$ to each spin at each time step $t>0$.
Stochastic
variables $\zeta_m (t)$ take on values 0 or 1 with equal probability.
For $k=2$ this and related linear dynamical rules lead to a
factorizable hierarchy
of relations for equal-time, $p$-spin correlation functions
[11,15]. The factorization property will also apply for $k>2$.

Essentially, the $p$-spin correlation functions
at fixed time $t$ can be
calculated from relations which only involve $p, (p-1), (p-2),
\ldots$-spin correlations, but do not couple to higher-order
correlations. The resulting expression are, however, difficult to treat
exactly, and all the available results for $k=2$ have been for 2-spin
correlations with symmetric (translationally-invariant) initial
conditions. In terms of the spin variables, the simplest tractable
initial conditions are fully random ones: spins $\sigma_j (0)$
are uncorrelated at different sites $j$, and distributed
identically at each $j$, with $\langle \sigma_j (0) \rangle = s$.
Here $s$ is a complex number which, for $k=3$, must
be within (or on the boundary of) the triangle defined by 1,
$\exp(2\pi i/3)$ and $\exp(4\pi i/3)$ in the complex plane.

The average $\langle \ldots \rangle$ for times $t>0$ will denote
both averaging over the stochastic evolution, i.e., over the choices
of the variables $\zeta_m (t)$, and over the distribution of the initial
values $\sigma_j (0)$. However, the $s\neq 0$ uncorrelated distribution
of the initial spin values implies a subtle correlation in the
initial distribution of the reacting-diffusing particles
on the dual lattice, as emphasized for $k=2$ in the Glauber-type
approach [9].

Thus, even though our solution can be easily extended to
general $s$, we will put $s=0$. In fact, we assume that the
reacting-diffusing particles are initially
distributed as follows for $k=3$:
at each dual-lattice site, one randomly places 0,
1, or 2 particles at time $t=0$. The average
initial density is $\rho (0)=1$,
and such distribution ensures that each spin has equal probability
to have the three allowed complex values (for $k=3$; extension
to $k\neq 3$ is trivial).

For times $t>0$ the particle density will decrease from $\rho(0)=1$.
In fact, the mean-field rate equation prediction for $k=3$ can be
easily derived, and it is $\rho (t) \sim 1/t $ for large times. We now
put $k=3$ and derive the exact expression for $\rho (t)$. The
result yields $\rho(t) \sim 1/\sqrt{t}$ for large times,
similar to the fluctuation-induced modification of the asymptotic
density in the case of the two-body annihilation $A+A \to {\rm
inert}$, which corresponds to another solvable case, $k=2$.

The case $k=3$ is special due to the property that the density of
particles on the dual lattice can be expressed
in terms of the 2-spin correlations of the $Z_k$ spins
for $k=3$ (and if fact also for $k=2$)
but not for $k>3$. Indeed, if $g=\exp
\left(2\pi i \ell /k \right)$ is one of the $k$ roots of
unity constituting the representation of $Z_{k>2}$, then
the appropriate phase number, $\ell=0,1,\ldots,k-1$, can
be calculated as a linear form $a_{(1)} + a_{(2)} g + a_{(3)} g^*$ only
for $k=3$. Indeed, three complex coefficients
$a_{(1),(2),(3)}$ are just the right
number to be fixed to get the three values $\ell=0,1,2$ properly.
For $k>3$, the expression required would involve higher integer
powers of $g$ and $g^*$ and it would no longer be linear.
For $k=3$ the explicit result is

$$ \ell = 1 + {1 \over \sqrt{3}} {\rm Re} \left[
(-\sqrt{3}+i) g\right] \eqno(3) $$

Recalling our rule of calculating the particle number $N_j(t)$,
we conclude that the average particle density
at site $j$ (of the dual lattice) at time $t$ is given by

$$ \rho (t) = 1 + {1 \over \sqrt{3}} {\rm Re} \left[
(-\sqrt{3}+i) \langle \sigma_{j+1} (t) \sigma^*_{j-1} (t)
\rangle \right] \qquad\qquad (k=3) \eqno(4) $$

\NI where we omitted the index $j$ for $\rho (t)$ due to translational
invariance. Relation (4) applies {\it only\/} for $k=3$
and expresses the particle density in terms of the 2-spin
correlation function of the $Z_3$ spins.

The dual-lattice sites can be empty, singly-occupied, or pair-occupied.
Let $\rho_{\rm single}  (t)$ denote the density
of those particles which are in
single-occupancy sites at time $t$. The density of particles which
are in pairs if then simply
$(\rho-\rho_{\rm single} )$. Note that the initial
conditions selected earlier correspond to
$\rho_{\rm single}  (0) = {1\over 3}$.
The single-occupancy property of a site is indicated by the value
of $\ell (2-\ell )=0,1,0$, for $\ell =0,1,2$, respectively.
This expression can also be written as a linear
form in $g$ and $g^*$ \ (for $k=3$). As a result we obtain the relation

$$ \rho_{\rm single}  (t) = {1\over 3} \left\{ 1 - {\rm Re} \left[
(1+i \sqrt{3}) \langle \sigma_{j+1} (t) \sigma^*_{j-1} (t)
\rangle \right] \right\} \qquad\qquad (k=3) \eqno(5) $$

Let us point out that all the ``linearity'' properties also apply for
$k=2$. In fact, all calculations are then with real variables: the
elements of $Z_2$ are simply $\pm 1$. The dynamics (2) for the case
$k=2$ describes particles hopping on a dual lattice and annihilating
pairwise on encounters, i.e., the cellular-automaton
(simultaneous-dynamics) variant of the reaction $A+A \to {\rm inert}$;
see [11]. Relation (4) is replaced by

$$ \rho (t) = {1\over 2} \left[ 1 -
\langle \sigma_{j+1} (t) \sigma_{j-1} (t)
\rangle \right]  \qquad\qquad (k=2) \eqno(6) $$

Turning back to $k=3$, we note that the dynamical rules (2) imply

$$ \langle \sigma_m (t) \rangle \equiv s =0 \eqno(7) $$

\noindent for translationally invariant initial conditions.
Since $s=0$, we can work with the \break
2-spin correlation function directly
(for $s\neq 0$ one has to consider the ``connected'' part; the
expressions are not much more complicated though). Thus, we define

$$ G_n (t) = \langle \sigma_j (t) \sigma^*_{j+n} (t)
\rangle \eqno(8) $$

\noindent which does not depend on $j$ due to translational invariance.
This correlation function can be considered for $n=0,2,4,\ldots$
only. Indeed, we have $G_{-n} (t) = G_n^*(t)$ generally.

By using (2), one can then derive the discrete-space and time diffusion
equation for $G_n (t)$,

$$ G_n (t+1) - G_n (t) = {1\over 4} \left[
G_{n+2} (t) - 2 G_n (t) + G_{n-2} (t) \right] \eqno(9) $$

\noindent which must be solved for $n=2,4,\ldots $ and $t=1,2,\ldots$,
subject to the initial and boundary conditions

$$ G_0 (t \geq 0) = 1 \qquad\;
{\rm and} \;\qquad G_{n>0} (t=0) =0 \eqno(10) $$

In writing (10), we used $s=0$ which implies a certain
simplification of the problem. Indeed, by (9) the correlation function
$G_n(t)$ is then {\it real\/} for these initial conditions. The form
of the solution is thus essentially identical to that for $k=2$.
Relations (4), (5), (6) yield, for {\it real\/} $k=3$ correlation
function,

$$ \left[ \rho (t) \right]_{k=3} =
2 \left[ \rho (t) \right]_{k=2} = 1-G_2 (t) \qquad\qquad (s=0)
\eqno(11) $$

$$ \rho_{\rm single}  (t) =
{1\over 3} \rho (t) \qquad\qquad (k=3,s=0) \eqno(12) $$

The discrete diffusion equation (9) can be analyzed be generating
function techniques [11]. The result for the density ($k=3$) is

$$ \rho (t) = { (2t+2) ! \over 2^{2t+1} \left[ (t+1) ! \right]^2 }
\eqno(13) $$

\noindent This function decreases smoothly from the initial value 1,
and in fact provides analytic continuation to all positive times $t$.
The large-time asymptotic form is

$$ \rho (t\gg 1) \simeq {2 \over \sqrt{\pi t}} \eqno(14) $$

The fact that the density for $k=3$ turns out nearly identical to that
for $k=2$ may at first seem disappointing. However, one must recall
that {\it all\/} the various exact calculations for model $1d$
reactions, based on the solution methods which involve evaluation of
2-spin correlation functions, always end up with continuous or
discrete versions of the diffusion equation,
--- property shared with the Glauber Ising model in $1d$. Thus, the
diversity of the functional forms obtained is limited [16].
The reaction scheme for $k=3$
is quite different from that for $k=2$. Indeed, if we denote
particles in singly-occupied sites by $A$, and the pairs of particles,
hopping as one unit, by $B$, then the reactions in the $k=3$ system are
$A+B \to {\rm inert}$, $A+A \to B$,\  $B+B \to A \,$. Generally, such
reaction scheme differs from the $k=2$ reaction, $A+A \to {\rm
inert}$, both in its mean-field rate equation form and in the time
dependence of densities and correlations.

Our results essentially suggest that for symmetric initial conditions
(i.e., equal concentrations ${1\over 3}$ of species $A$ and $B$), the
densities of the $k=3$ and $k=2$ reaction schemes are proportional, and
both are fluctuation-dominated, $\sim 1/\sqrt{t}$, as opposed to the
rate-equation prediction, $\sim 1/t$, for large times.

Helpful comments by Dr.~E.G.~Klepfish and Dr.~D.A.~Rabson
are greatly appreciated.
This research was partially supported by the
Science and Engineering Research Council (UK)
under grant number GR/G02741.
The author also wishes to acknowledge
the award of a Guest Research Fellowship at Oxford from
the Royal Society.

\NP

\centerline{\bf REFERENCES}

{\frenchspacing

\item{1.} M. Bramson and D. Griffeath,
Ann. Prob. {\bf 8}, 183 (1980).

\item{2.} D.C. Torney and H.M. McConnell,
J. Phys. Chem. {\bf 87}, 1941 (1983).

\item{3.} K. Kang, P. Meakin, J.H. Oh and S. Redner,
J. Phys. A{\bf 17}, L665 (1984).

\item{4.} T. Liggett, {\sl Interacting Particle Systems\/}
(Springer-Verlag, New York, 1985).

\item{5.} Z. Racz, Phys. Rev. Lett. {\bf 55}, 1707 (1985).

\item{6.} A.A. Lushnikov, Phys. Lett. A{\bf 120}, 135 (1987).

\item{7.} M. Bramson and J.L. Lebowitz, Phys. Rev. Lett. {\bf 61},
2397 (1988).

\item{8.} Review: V. Kuzovkov and E. Kotomin, Rep. Prog. Phys.
{\bf 51}, 1479 (1988).

\item{9.} J.G. Amar and F. Family, Phys. Rev. A{\bf 41}, 3258 (1990).

\item{10.} L. Braunstein, H.O. Martin, M.D. Grynberg
and H.E. Roman, J. Phys. A{\bf 25}, L255 (1992).

\item{11.} V. Privman, J. Stat. Phys. (1992), in print.

\item{12.} D. ben--Avraham, M.A. Burschka and C.R. Doering,
J. Stat. Phys. {\bf 60}, 695 (1990), and references therein.

\item{13.} A.J. Bray, J. Phys. A{\bf 23}, L67 (1990).

\item{14.} B. Derrida, C. Godr\`eche and I. Yekutieli,
Phys. Rev. A{\bf 44}, 6241 (1992).

\item{15.} M. Scheucher and H. Spohn, J. Stat. Phys.
{\bf 53}, 279 (1988), and references therein.

\item{16.} Dynamical problems in $1d$ can also be reformulated as
eigenvalue problems of certain quantum spin systems.
\ M. Droz, M. Henkel and V. Rittenberg, private communication (1992),
have explored this connection, which should allow the use of
the Baxter-Yang equation approach to identify new integrable systems
without the restriction to special cases such as relations (6) or (4).

}

\bye